\documentclass[prd,twocolumn,superscriptaddress,showpacs,nofootinbib,preprintnumbers]{revtex4}

\usepackage{amsmath}
\usepackage{amsfonts}
\usepackage{graphicx}
\usepackage{dcolumn}

\def\be{\begin{equation}}
\def\ee{\end{equation}}
\def\ba{\begin{eqnarray}}
\def\ea{\end{eqnarray}}
\def\bs{\begin{subequations}}
\def\es{\end{subequations}}

\newcommand{\p}{{\partial}}

\def\f{\frac}
\newcommand{\hm}{{\cal C}_{\mathrm{M}}}
\def\R{\mathbb{R}}
\newcommand{\heff}{{\cal C}_{\mathrm{eff}}}
\usepackage{color}

\def\c{\mathfrak{c}}
\def\p{\mathfrak{p}}

\begin{document}

\title{Avoidance of future singularities in loop quantum cosmology}

\bigskip

\author{M.~Sami}
\affiliation{Centre for Theoretical Physics, Jamia Millia, New
Delhi-110025, India}

\affiliation{Department of Physics, Jamia Millia,
New Delhi-110025, India}
\email{sami@jamia-physics.net}

\author{Parampreet Singh}
\affiliation{Institute for Gravitational Physics and Geometry,
Physics Department, Penn State, University Park, PA 16802, U.S.A.}
\email{singh@gravity.psu.edu}

\author{Shinji Tsujikawa}
\affiliation{Department of Physics, Gunma National College of
Technology, Gunma 371-8530, Japan}
\email{shinji@nat.gunma-ct.ac.jp}

\date{\today}

\preprint{IGPG-06/5-4}

\begin{abstract}

We consider the fate of future singularities in the effective
dynamics of loop quantum cosmology. Non-perturbative
quantum geometric effects which lead to $\rho^2$ modification
of the Friedmann equation at high energies result in generic
resolution of singularities whenever energy density $\rho$
diverges at future singularities of Friedmann dynamics.
Such quantum effects lead to the avoidance of
a Big Rip, which is followed by a
recollapsing universe stable against perturbations.
Resolution of sudden singularity, the case
when pressure diverges but energy density approaches a finite
value depends on the ratio of the latter to a critical energy
density of the order of Planck. If the value of this ratio is greater
than unity, the universe escapes the sudden
future singularity and becomes oscillatory.

\end{abstract}


\maketitle

\section{Introduction}

Observations suggest that the current universe is dominated
by a matter component which leads to an accelerated expansion
of the universe-- called dark energy (see Refs.~\cite{review}
for review). This has stimulated a study that our universe
may face a future singularity.
Such future singularities typically arise if the universe
is dominated by matter which violates dominant energy condition and
causes a state of super-acceleration of the universe before leading it
to a singularity. They can occur due to divergence
either of the energy density $\rho$ and/or the pressure density
$p$ of the matter content.
For example if the universe is filled with a phantom dark energy
with a constant equation of state $w$ less than
$-1$ \cite{Caldwell}, this leads to a Big Rip singularity
at which both $\rho$ and $p$ diverge with a finite
time \cite{CKW}.
Barrow pointed out a possibility to obtain a sudden future
singularity at which $\rho$ is finite but
$p$ diverges \cite{Barrow}.
Depending on the equation of state of dark energy, future
singularities have been categorized in different
classes \cite{NOT}.

Existence of future singularities in Friedmann-Robertson-Walker (FRW)
cosmology reflects the vulnerability of standard Friedmann dynamics
whenever $\rho$ or $p$ become of the order of Planck values.
This indicates that limit of validity of general relativity
has been reached and inputs from quantum gravity are necessary to
probe the dynamics near the singularity. Resolution of
singularities using Wheeler-DeWitt
quantization has been attempted \cite{WD} but has met with little success. 
One of the primary
reasons for its failure has been a lack of a fundamental theory which
can guide quantization in the Wheeler-DeWitt framework. Issue of
resolution of past \cite{string} and future singularities \cite{sttt}
have been investigated using perturbative corrections in string theoretic
models. These analysis indicate that generic resolution of
singularities may only be accomplished using non-perturbative
corrections. In particular in the absence of an analysis which uses
non-perturbative quantum gravitational modifications to model the
dynamics of dark energy, the fate of future singularities has remained
an open problem.

Loop quantum gravity (LQG) is a leading background independent
non-perturbative quantization of gravity \cite{lqg_review}
which has been very well understood in the cosmological setting in
loop quantum cosmology (LQC) \cite{mblr}. To its success, LQG has dealt with
various singularities in cosmological setting
\cite{sing,APS1,APS2,APS3,svv} and techniques have also
been used to resolve singularities in black hole
spacetimes \cite{sing1}.  Recent investigations have revealed that
non-perturbative loop
quantum effects lead to a $\rho^2$ modification of the Friedmann
equation with a negative sign \cite{APS2,singh:2006a,abhay}. The
modification becomes important when energy density of the universe
becomes of the same order of a critical density $\rho_c$.
The resulting dynamics generically leads to a bounce when our flat
expanding universe is evolved backwards \cite{APS1,APS2,APS3,svv}.

Since important insights have been gained on resolution of
space-like singularities in LQC, it offers a natural arena
to investigate the fate of future singularities.
This is the goal of the present work. Using the
effective Friedmann dynamics which has emerged from LQC
we would analyze the way non-perturbative quantum
gravitational effects modify the dynamics near future singularities.
The plan of this paper is as follows. In the next section we would
briefly review the way an effective modified Friedmann dynamics
is obtained from the discrete quantum dynamics in LQC. Since,
Wheeler-DeWitt quantization, which like LQC is a mini-superspace
approach, has been unsuccessful in resolving
space-like singularities we would highlight some differences which
emerge with LQC (for details see Ref.\,\cite{APS2}). In Sec. III we
analyze in detail the fate of three types of future singularities--
type I (the Big Rip): scale factor $a$, energy density $\rho$
and pressure $p$ becoming infinite in
finite time, type II (sudden): $p$ becoming infinite with finite $\rho$
in finite time, and type III: $\rho$ and $p$ diverging with finite $a$
in finite time.
We will show the LQC can successfully resolve
type I and type III singularities for generic choice of initial
conditions. Resolution of type II singularities though depends on the
amplitude of model parameters. We  conclude with a
summary of our results in Sec. IV.

\section{Effective dynamics in Loop Quantum Cosmology}

In LQG the phase space of classical general relativity is expressed in terms of SU(2)
connection $A_a^i$ and densitized triads $E^a_i$. In loop quantum
cosmology (LQC) \cite{mblr}, due to underlying symmetries of the FRW spacetime the
phase space structure simplifies and can be casted in terms of
canonically conjugate connection $\c$ and triad $\p$ which satisfy
$\{\mathfrak{c},\mathfrak{p}\} = \kappa \gamma/3$, 
where $\kappa=8\pi G$ ($G$ is 
gravitational constant) and $\gamma$ is the dimensionless
Barbero-Immirzi parameter (which is set by the black hole
thermodynamics in LQG, as $\gamma \approx 0.2375$).
On the space of
physical solutions of general relativity they are related to scale
factor and its time derivative as: $\c = \gamma \dot a$ and $\p = a^2$.

The elementary
variables used for quantization in LQC are the triads and
holonomies of connection over
edges of loops: $h_i(\mu) = \cos(\mu \c/2) + 2 \sin(\mu \c/2) \tau_i$,
where $\tau_i$ are related to Pauli spin matrices as $\tau_i = - i
\sigma_i/2$ and $\mu$ is related to  the length of the edge over which
holonomy is evaluated.
The algebra generated by holonomies is that of almost periodic functions of $\c$ with elements of the form:
$\exp(i \mu \c/2)$. On quantization, though holonomies have well defined
quantum operators there are no quantum operators for $\c$ in
LQC (as in LQG). The kinematical Hilbert space in LQC is ${\cal
H} = L^2(\R_{\mathrm{Bohr}},d\mu_{\mathrm{Bohr}})$
where $\R_{\mathrm{Bohr}}$ is the Bohr compactification of the real line
and $\mu_{\mathrm{Bohr}}$ is the Haar measure on it. Note that the
Hilbert space is different from the one in Wheeler-DeWitt
quantization: ${\cal
H}_{\mathrm{WDW}} = L^2(\R,d \mu)$. The triad and thus the scale
factor operator in LQC have a discrete eigenvalue spectrum and quantum
constraint, obtained by expressing the classical constraint in terms
of holonomies and positive powers of triad and then quantized, in LQC
leads to a discrete quantum difference equation whose {\it all}
solutions are non-singular -- another important distinction from the
Wheeler-DeWitt theory.

Physical predictions can be extracted from LQC by construction of a
physical Hilbert space. By identifying Dirac observables on this space,
information about dynamics can be extracted using ideas of emergent
time. On constructing coherent states we can then find out the
expectation values of Dirac observables and compare the quantum
dynamics with the classical one. It turns out that when a flat expanding
universe is evolved backward using loop quantum dynamics, instead of
ending in  big bang singularity it bounces at Planck scale to a
contracting branch \cite{APS1,APS2,APS3}.

The coherent states used to analyze the details of quantum dynamics
also play an important role in obtaining an effective Hamiltonian
description of dynamics governed by quantum difference equation. This
can be done by using methods of geometric formulation of quantum
mechanics \cite{as} where one notes that quantum Hilbert space can be
regarded as a quantum phase space with a bundle structure. The
classical phase space forms the base of this bundle, whereas fibers
consist of states with same expectation values of conjugate
variables. Horizontal sections
of the bundle are isomorphic to the classical phase space. Using
coherent states one can then find horizontal sections which are
preserved by quantum evolution which then leads us to an effective
Hamiltonian with loop quantum modifications \cite{josh,sv}:
\be
\heff = - \f{3}{\kappa \gamma^2 \bar \mu^2}  \,a \sin^2(\bar \mu \c) +
\hm ~.
\ee
Here $\bar \mu$ is the kinematical length of the edge of a square loop
which has the area given by the minimum eigenvalue of the area
operator in LQG \cite{abhay} and $\hm$ corresponds to matter Hamiltonian which
in general contains
modifications due to regularization of inverse scale factor \cite{mblr}. These
modifications are negligible for large universes and would not be
considered in the present work.

The modified Friedmann equation can then be obtained by using the
Hamilton's equation:
\be
\dot \p = \{\p,\heff\} = - \f{\kappa \gamma}{3}
\f{\partial \heff}{\partial \c}
\ee
in the effective Hamiltonian constraint
$\heff \approx 0$:
\be\label{basic1}
H^2=\frac{\kappa}{3} \rho
\left(1-\frac{\rho}{\rho_{c}}\right)\,,
\ee
with $\rho_c = \sqrt{3}/(16 \pi \gamma^3 G^2 \hbar)$ \cite{abhay}
where $\hbar$ is Planck constant.
Along with the conservation law:
\be\label{basic2}
\dot{\rho}+3H(\rho+p)=0\,,
\ee
Eq.~(\ref{basic1}) provides an effective description of Friedmann
dynamics which very well approximates the underlying discrete quantum
dynamics and confirms with the picture of bounce which occurs when
$\rho = \rho_c$ (of the order of Planck density)
\cite{APS2,APS3,svv}. In the classical limit $\hbar \rightarrow 0$
one has $\rho_c \rightarrow \infty$, thus classically non-singular bounce is
absent. Further, for $\rho \ll \rho_c$ the modified Friedmann equation
reduces to the standard one. Interestingly $\rho^2$ modifications also appear in string
inspired braneworld
scenarios and it turns out that there exist interesting dualities
between two frameworks \cite{singh:2006a}. However, such modifications
in braneworlds usually appear with a positive sign and a bounce is
absent (unless one assumes existence of two time-like dimensions \cite{YS}).
We will now address the issue of future singularities in the effective
dynamics of LQC. Our treatment of matter which leads to such
singularities would be phenomenological and at an effective level (as
in Ref. \cite{singh:2005a}).

\section{Avoidance of future singularities}

For the analysis  of the fate of future singularities in LQC, it is
useful to first obtain the rate of change of Hubble parameter from
Eqs.~(\ref{basic1}) and (\ref{basic2}):
\be
\label{basic3}
\dot{H}=-\frac{\kappa}{2}(1+w)
\rho \left(1-2 \frac{\rho}{\rho_{c}} \right)\,,
\ee
where $w$ is the equation of state: $w = p/\rho$ which in general may
not be a constant.

It is convenient to define two variables
\be
\label{xdef}
x \equiv \frac{\kappa \rho}{3H^2}\,, \quad
y \equiv \frac{\rho}{\rho_{c}}\,.
\ee
Then from Eq.~(\ref{basic1}) we find
\be
\label{yx}
y=1-1/x\,.
\ee
Since $H^2$ is positive, the variables $y$ and $x$ are
in the ranges $0<y<1$ and $x>1$.
{}From Eqs.~(\ref{basic2}) and (\ref{basic3})
we obtain the differential equation for the variable $x$:
\be
\label{difeq}
\frac{{\rm d}x}{{\rm d}N}=-3(1+w)x (x-1)\,,
\ee
where $N \equiv {\rm ln}\, (a)$.

Since $H$ can change the sign, it will be convenient to solve
differential equations in terms of a cosmic time $t$
rather than $N$.
Defining two dimensionless quantities $\tilde{t} \equiv H_{c}t$
and $\tilde{H} \equiv H/H_{c}$, where
$H_{c} \equiv \sqrt{\kappa
\rho_{c}/3}$, we find that Eqs.~(\ref{basic2}) and (\ref{basic3})
are written as
\ba
\label{nu1}
& &\frac{{\rm d}y}{{\rm d}\tilde{t}}=-3(1+w)\tilde{H}y\,,\\
\label{nu2}
& &\frac{{\rm d}\tilde{H}}{{\rm d}\tilde{t}}=-\frac32
(1+w)y(1-2y)\,,
\ea
together with the constraint equation
\ba
\label{Hdl}
 \tilde{H}^2=y(1-y)\,.
\ea
Combining equations (\ref{nu1}) and (\ref{nu2}) gives
\ba
\label{sorder}
\frac{{\rm d}^2 y}{{\rm d} \tilde{t}^2}=\frac92
(1+w)^2y^2 (3-4y)-3\tilde{H}y \frac{{\rm d}w}
{{\rm d}\tilde{t}}\,.
\ea

We will study several equations of state which, in standard
Einstein gravity, give rise to
various types of future singularities \cite{NOT}.
Our interest is to clarify the role of loop quantum
modifications on the following singularities
which are known to exist in standard Einstein gravity:
\begin{itemize}
\item  Type I (``Big Rip'') : For $t \to t_s$,
$\rho \to \infty$, $|p| \to \infty$, $H \to \infty$ and
$a \to \infty$
\item  Type II (``sudden'') : For $t \to t_s$
$\rho \to \rho_s$, $|p| \to \infty$,
$H \to H_{s}$ and $a \to a_s$
\item  Type III : For $t \to t_s$,
$\rho \to \infty$, $|p| \to \infty$,
$H \to \infty$ and $a \to a_s$.
\end{itemize}
Here $t_s$, $\rho_s$, $H_{s}$ and $a_{s}$ are constants.
The type I singularity appears for constant $w$
less than $-1$ \cite{CKW}.
The type II is a sudden future
singularity \cite{Barrow} at which
$\rho$ and $a$ are finite but $p$ diverges.
The type III appears for the model with $p=-\rho-A \rho^{\beta}$
with $\beta>1$ \cite{Stefancic}.
In what follows we shall study each case separately.

\subsection{Type I singularity}

Let us consider a constant equation of state, $w$.
In this case Eq.~(\ref{difeq}) is easily integrated to give
\be
x=\frac{1}{1-Aa^{-3(1+w)}}\,, \quad
y=Aa^{-3(1+w)}\,,
\ee
where $A$ is a positive constant.

When $w>-1$ the solutions in an expanding universe approach the
fixed point $(x, y)=(1, 0)$, which corresponds to the standard
Einstein gravity. Meanwhile when $w<-1$ one has $x \to \infty$ and
$y \to 1$ with a scale factor satisfying $Aa_c^{-3(1+w)}=1$. In this
case $\rho$ approaches a constant value $\rho_c$ as $a \to a_c$.
{}From Eq.~(\ref{basic1}) we find that the Hubble parameter becomes
zero at this point. This equation [or equivalently Eq.~(\ref{Hdl})]
also tells us that the Hubble parameter $\tilde{H}$
varies between its maximum and minimum values
given by $\tilde{H}_{\rm max}=1/2$ and $\tilde{H}_{\rm min}=-1/2$
respectively. We also notice from Eq.~(\ref{nu2}) that
${\rm d}\tilde{H}/{\rm d}\tilde{t}<0$
at the time when $y$ becomes $y=1$,  leading to the decrease of $\tilde{H}$.
Equation (\ref{nu1}) tell us that $y$ begins to decrease
after it has reached its maximum value $y=1$ corresponding to $\tilde{H}=0$.

We can now understand the fate of the universe for $w<-1$
in LQC. A qualitative description of the evolution can be obtained by
using Eqs.~(\ref{nu1}), (\ref{nu2}) and (\ref{Hdl}).
Let us begin to examine the evolution with a positive
initial value of $\tilde{H}$. {}From Eqs.(\ref{nu1})
and (\ref{nu2}), we find that both $y$ and $\tilde{H}$
grow until $y$ reaches $y=1/2$. Such
a behavior of $\rho$ and $H$ is generic to a phantom
dominated universe with constant $w$ in the standard FRW cosmology
which then leads to a Big Rip.
The LQC correction changes this cosmic evolution
in a crucial manner.
The Hubble rate begins to decrease after it reaches a maximum value
at $y=1/2$, whereas $y$ continues to grow
until $\tilde{H}$ drops below zero [see Eqs.~(\ref{nu1}) and (\ref{nu2})].
As explained above, the variable $y$ starts to decrease with
${\rm d}\tilde{H}/{\rm d}\tilde{t}<0$
after it has reached its maximum value $y=1$.
When $y$ becomes smaller than $1/2$, ${\rm d}\tilde{H}/{\rm d}\tilde{t}$
changes its sign after which $\tilde{H} (<0)$ increases toward 0.
This stage corresponds to a recollapsing universe that asymptotically
approaches the fixed point $(x,y)=(1,0)$.
{}From Eq.~(\ref{sorder}) together with Eq.~(\ref{Hdl}) we find that the
asymptotic behavior is given by
$y \propto t^{-2}$ and $H \propto -t^{-1}$.
Note that for negative $\tilde{H}$ the fixed point $(x,y)=(1,0)$ is
stable against perturbations as can be checked
by linearly perturbing the system (\ref{nu1}) and (\ref{nu2}).
In fact two eigenvalues of a matrix for perturbations \cite{CLW}
are $0$ and $-3(1+w)\tilde{H}$, where the latter
is negative for $\tilde{H}<0$.

\begin{figure}
\includegraphics[height=2.9in,width=3.4in]{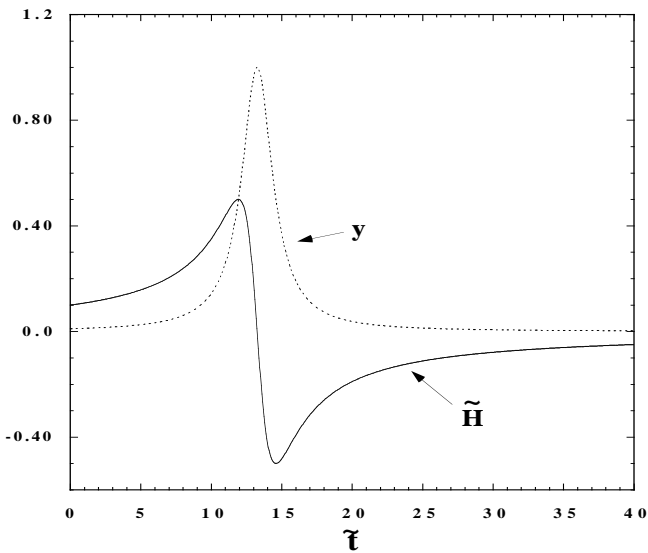}
\caption{Evolution of the Hubble parameter and the variable
$y=\rho/\rho_{c}$ for $w=-1.5$ with initial conditions
$y_{i}=0.01$ and $\tilde{H}_{i}=[y_i(1-y_{i})]^{1/2}$.
The Big Rip singularity is avoided in the presence of
loop quantum modifications to the Friedmann dynamics.
}
\label{fig1}
\end{figure}

We have numerically solved Eqs.~(\ref{nu1}) and (\ref{nu2})
for $w<-1$ with initial conditions $\tilde{H}>0$.
In Fig.~\ref{fig1} we plot an example for
the evolution of $\tilde{H}$ and $y$ when $w=-1.5$.
Our numerical results clearly
confirm the qualitative behavior of the evolution presented above.

Thus we have shown the Big Rip singularity
is beautifully avoided in the frame work of LQC.
The solutions finally approach a contracting universe
in standard Einstein gravity ($H \to 0$ and $\rho \to 0$
as $t \to \infty$).
We note that when $w>-1$ bouncing solutions can be
obtained if $H<0$ initially \cite{APS1,APS2,APS3,svv}.

\subsection{Type II singularity}

In standard Einstein gravity the type II singularity appears
when the pressure density $p$ diverges as $\rho$ approaches
some constant value $\rho_{0}$.
For example, this is realized when $p$ is given
by \cite{NOT}
\be
\label{pexp}
p=-\rho-\frac{B}{(\rho_{0}-\rho)^{\gamma}}\,,
\ee
where $B$, $\rho_{0}$ and $\gamma$ are positive
constants. This singularity appears at a finite time
as $\rho$ approaches $\rho_{0}$.

Let us consider the cosmological dynamics in the presence of
the loop correction.
The equation of state is now dependent on $\rho$, i.e.,
$w=-1-B/\rho(\rho_{0}-\rho)^{\gamma}$.
Substituting this expression for Eq.~(\ref{difeq})
by using the relation (\ref{yx}), we get
\be
\frac{{\rm d}x}{{\rm d}N}
=\tilde{B} \frac{x^2}{(r-1+1/x)^{\gamma}}\,,
\ee
where $\tilde{B} \equiv 3B/\rho_{c}^{1+\gamma}$ and
$r \equiv \rho_{0}/\rho_{c}$\,.
Integrating this equation gives
\be
\left( r-1+\frac{1}{x} \right)^{\gamma+1}=
\left( r-1+\frac{1}{x_{i}} \right)^{\gamma+1}
-\tilde{B} (\gamma +1)N\,,
\ee
where we chose the initial condition $x=x_{i}$ at $N=0$.
This shows that $x$ gets larger with the increase of $N$,
in which case $y=\rho/\rho_{c}$ grows from Eq.~(\ref{yx}).
The solutions approach $x \to \infty$ and $y \to 1$
provided that $\rho$ does not pass the singularity at
$\rho=\rho_{0}$ before reaching $\rho=\rho_{c}$.

\begin{figure}
\includegraphics[height=2.8in,width=3.4in]{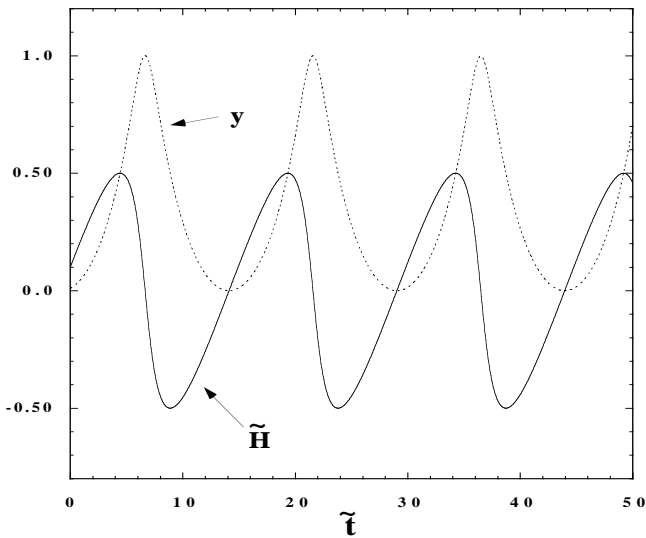}
\caption{Evolution of the Hubble parameter and the variable
$y=\rho/\rho_{c}$ for the model (\ref{pexp}) with
$\tilde{B}=1$, $r=\rho_0/\rho_{c}=2$ and $\gamma=2$.
We choose initial conditions
$y_{i}=0.01$ and $\tilde{H}_{i}=[y_i(1-y_{i})]^{1/2}$.
The Hubble parameter $\tilde{H}$ oscillates between $-1/2$ and $1/2$
without reaching the singularity at $\rho=\rho_{0}$.
}
\label{fig2}
\end{figure}

When $\rho_{0} > \rho_{c}$ the system reaches
$\rho=\rho_{c}$ with a finite time $N_{c}$
satisfying $\tilde{B}(\gamma+1)N_c
=(r-1+1/x_i)^{\gamma+1}-(r-1)^{\gamma +1}$.
The Hubble parameter vanishes at this point, since
$H^2=\kappa \rho/3x$ from Eq.~(\ref{xdef}).
{}From Eqs.~(\ref{nu1}) and (\ref{nu2}) the differential equations
for $y$ and $H$ are given by
\be
\frac{{\rm d}y}{{\rm d}\tilde{t}}=
\frac{\tilde{B}}{(r-y)^\gamma} \tilde{H}\,, \quad
\frac{{\rm d}\tilde{H}}{{\rm d}\tilde{t}}=
\frac{\tilde{B}(1-2y)}{2(r-y)^\gamma}\,.
\label{typeII}
\ee
When $y=1$ one has ${\rm d}y/{\rm d}\tilde{t}=0$
and ${\rm d}\tilde{H}/{\rm d}\tilde{t}<0$.
Then the Hubble parameter becomes negative,
which is accompanied by the decrease of $y$.
{}From Eq.~(\ref{typeII}) we find
${\rm d}\tilde{H}/{\rm d}\tilde{t}>0$
for $y<1/2$, during which $\tilde{H}$ increases.
In the type I case $y$ and $\tilde{H}$ asymptotically approach zero
with time-dependence $y \propto t^{-2}$ and $H \propto -t^{-1}$.
The type II case is different because of a time-dependent equation of
state. In fact when $\tilde{H}=0$ and $y=0$  we find
${\rm d}y/{\rm d}\tilde{t}=0$
and ${\rm d}\tilde{H}/{\rm d}\tilde{t}>0$.
This behavior is clearly seen in Fig.~\ref{fig2}.
Both $\tilde{H}$ and $y$ increase after the system passes
the point $\tilde{H}=0$ and $y=0$, which is followed by
the maximum value of $\tilde{H}$ at $y=1/2$.
After that the evolution of the universe mimics the previous one,
namely $\tilde{H}$ oscillates between $-1/2$ and $1/2$
together with the oscillation of $y$ between 0 and 1.
Hence the universe repeats the cycle of expansion and contraction
without reaching any singularities (see Fig.~\ref{fig2}).

When $\rho_0<\rho_{c}$, independently of $\rho>\rho_0$ or $\rho<\rho_0$, the
solutions reach the sudden future singularity at $\rho=\rho_{0}$
in a finite time.

%
\subsection{Type III singularity}

The type III singularity appears for the model
\be
\label{typeIII}
p=-\rho- C \rho^{\beta}\,, \quad \beta>1\,,
\ee
where $C$ is a positive constant.
Integrating Eq.~(\ref{basic2}) for the equation of
state (\ref{typeIII}), we find that the scale factor
is given by
\be
a=a_{0} \exp \left( \frac{\rho^{1-\beta}}
{3C(1-\beta)} \right)\,,
\ee
where $a_{0}$ is a constant.
In Einstein gravity one has $\rho \to \infty$ and
$|p| \to \infty$ in
a finite time, but $a$ is finite when $\beta>1$.
Hence this is different from the Big Rip singularity
at which scale factor diverges.

In LQC Eq.~(\ref{difeq}) gives
\be
\frac{{\rm d} x}{{\rm d}N}=
\tilde{C} x^2 \left( 1-\frac{1}{x} \right)^{\beta}\,,
\ee
where $\tilde{C} \equiv 3C \rho_{c}^{\beta -1}$.
This is integrated as
\be
\left( \frac{1}{1-1/x} \right)^{\beta-1}=
\left( \frac{1}{1-1/x_{i}} \right)^{\beta-1}-
\tilde{C} (\beta-1)N\,.
\ee
Then we get $x \to \infty$, $y \to 1$ and $H \to 0$ as $N \to N_c$,
where $N_{c}$ is given by
$\tilde{C}(\beta -1)N_c=(1/(1-1/x_{i}))^{\beta -1}-1$.
{}The differential equations (\ref{nu1}) and (\ref{nu2}) are
\be
\frac{{\rm d}y}{{\rm d}\tilde{t}}=
\tilde{C}\tilde{H}y^\beta\,, \quad
\frac{{\rm d}\tilde{H}}{{\rm d}\tilde{t}}=
\frac{\tilde{C}}{2}y^{\beta} (1-2y)\,.
\label{typeIIIeq}
\ee
Hence one has ${\rm d}y/{\rm d}\tilde{t}=0$
and ${\rm d}\tilde{H}/{\rm d}\tilde{t}<0$ for $y=1$ and $\tilde{H}=0$,
which is followed by the decrease of $y$ and $\tilde{H}$ ($<0$).
The evolution of the system is similar to what we discussed
in the type I case. After the Hubble rate reaches a minimum at $y=1/2$,
$\tilde{H}$ and $y$ asymptotically approach $\tilde{H}=y=0$.
When $y \ll 1$, in fact, we have
${\rm d}^2 y/{\rm d}\tilde{t}^2 \approx
(3\tilde{C}^2/2)y^{2\beta}$ from Eq.~(\ref{sorder}), which gives
$y \propto t^{-2/(2\beta-1)}$ and $\tilde{H} \propto  -t^{-1/(2\beta-1)}$.
Hence the final attractor is a contracting universe with
$\rho \to 0$, $p \to 0$ and $H \to -0$ as $t \to \infty$.

\section{Conclusions}

In this paper we have studied the avoidance of future singularities
using the effective dynamics of loop quantum cosmology.
Non-perturbative quantum effects give rise to a $\rho^2$ correction
whose effect depends upon the ratio $\rho/\rho_{c}$, where
$\rho_{c}$ is of order of Planck density.
Typically this type of correction is thought to be important only in
early universe whose energy density is close to $\rho_{c}$, but
it can be also important in future universe if (phantom) dark energy
is present as observations suggest.
Note that the modifications we studied are different from those given
which emerge from the regularization of
inverse scale factor operator in LQC and can be important below
a critical scale factor $a_*$ (see for e.g., Ref.~\cite{original}).
These corrections are negligible for $a
\gg a_*$ and are not considered in the present work
which deals with late time expansion dynamics in LQC.

There are several types of future singularities which appear in
standard Einstein gravity. In the type I case where $\rho$, $p$
and $a$ diverge in a finite time and in the type III case where
$\rho$ and $p$ are infinite but $a$ is finite in a finite time,
we find that the loop quantum modifications generically
remove these singularities.
The universe transits from an expanding branch to a contracting
branch after the energy density approaches critical value $\rho_{c}$.
After the Hubble parameter reaches a negative minimum when
$\rho/\rho_{c}=1/2$, it increases toward a stable fixed
point $H=0$ in an infinite time (see Fig.~\ref{fig1}). The fate of the
universe thus dramatically changes on considering loop quantum
modifications in the standard Friedmann dynamics.

In the type II case where $p$ diverges but $\rho$ ($\rho_{0}$)
and $a$ are finite in a finite time,
sudden singularity is not removed
when $\rho_{0}$ is smaller than $\rho_{c}$.
When $\rho_{0}>\rho_{c}$, however,
the Hubble parameter $H$ exhibits an oscillation around
$H=0$ (see Fig.~\ref{fig2}). This corresponds to
an oscillating universe without any singularities.

We have thus shown that in most cases the future
singularities are avoided because of the presence of
loop quantum corrections. Our analysis of the resolution of
singularities clearly reflects the important role played by
non-perturbative quantum gravity modifications in order to
fully understand the dynamics of universe around
the Planck energy.

\section*{ACKNOWLEDGMENTS} PS thanks Kevin Vandersloot for useful discussions. PS is supported by  by NSF grants
PHY-0354932 and PHY-0456913 and the Eberly research funds of Penn
State. ST is supported by JSPS (Grant No.\,30318802). MS thanks
IUCAA for hospitality.


\end{document}